# Experimental and theoretical analyses of strongly polarized photon emission from non-polar InGaN quantum dots

**Tong Wang[1], Tim J. Puchtler[1], Saroj Kanta Patra[2,3], Tongtong Zhu[4], Muhammad Ali[5], Tom Badcock[5], Tao Ding[4,6], Rachel A. Oliver[4], Stefan Schulz[2] and Robert A. Taylor[1]**

[1]Department of Physics, University of Oxford, Parks Road, Oxford, OX1 3PU, UK

[2]Tyndall National Institute, University College Cork, Cork, Ireland

[3]Department of Electrical Engineering, University College Cork, Ireland

[4]Department of Materials Science and Metallurgy, University of Cambridge, Cambridge, CB3 0FS, UK

[5]Cambridge Research Laboratory, Toshiba Research Europe Limited, Cambridge, CB4 0FS, UK

[6]Nanophotonics Centre, Cavendish Laboratory, University of Cambridge, Cambridge, CB3 0HE, UK

Email: tong.wang@physics.ox.ac.uk; sarojkanta.patra@tyndall.ie

**Abstract**

We present a comprehensive investigation of the polarization properties of non-polar *a*-plane InGaN quantum dots (QDs) and their origin with statistically significant experimental data and rigorous **k p** modelling. The unbiased selection and study of 180 individual QDs allow us to compute an average polarization degree of 0.90, with a standard deviation of only 0.08. When coupled with theoretical insights, we show that *a*-plane InGaN QDs are highly insensitive to size differences, shape anisotropies, and indium content fluctuations. Furthermore, 91% of the studied QDs exhibit a polarization axis along the crystal [1-100] axis, with the other 9% polarized orthogonal to this direction. When coupled with their ability to emit single-photons, *a*-plane QDs are good candidates for the generation of linearly polarized single-photons, a feature attractive for quantum cryptography protocols.

## 1. Introduction

On-demand polarized single-photon sources [1,2] are essential for applications in quantum information sciences, such as quantum key distribution [3–5] and optical quantum computing [6]. To develop these non-classical light sources, semiconductor quantum dots (QDs) [7–9] have been a field of nanotechnology under intense scientific investigation. Although ultrapure and highly indistinguishable single-photon generation has been achieved in various arsenide-based QD systems [10–14], the large band offsets and strong exciton binding energies of III-nitride materials are needed for the realization of polarized photon emission [15–17] and room-temperature operation [18,19]. These polarized single-photon sources can then fulfil the need for on-chip polarization encoding in quantum cryptography, such as the BB84 protocol [20]. However, previous studies of polar *c*-plane (0001) InGaN QDs reveal that only selected QDs with large shape anisotropies can exhibit high degrees of optical linear polarization (DOLP) [15,16], which are also in agreement with our own investigations [21]. Furthermore, in most *c*-plane nitride QD systems, different anisotropies could result in random polarization axes, which are highly undesirable in the implementation of polarization-based protocols in potential applications. Consequently, much effort is needed to define the anisotropy, in order to obtain consistently high DOLPs with deterministic polarization directions. The only such experimental report in the nitrides is on the manipulation of pyramidal geometries [22], where the geometry of the structure makes electrical contacting challenging. As such, a simpler and more practical method is needed to achieve polarization control, for the realization of on-chip polarized single-photon generation and related applications.

In this work, we report a combined theoretical and experimental investigation of a planar self-assembled semiconductor QD system – non-polar *a*-plane (11-20) InGaN/GaN QDs – that provides not only a statistically high DOLP but also a fixed polarization direction determined by the underlying material crystallography. Furthermore, while the DOLPs of *a*-plane quantum wells (QWs) are limited to ~ 0.60 [23], we report a much higher average DOLP of ~ 0.90 in *a*-plane InGaN QDs. Previous investigations have reported polarization properties of some *a*-plane InGaN QDs along [0001] [24], orthogonal to those reported in *a*-plane QWs (along [1-100]) [23,25,26]. We find that this is a special case to the general behaviour of *a*-plane QDs. Here, more rigorous measurements of the DOLP have been made upon a large number of QDs using high extraction efficiency QD-nanopillar structures.

Theoretical simulations have also been undertaken to elucidate the origin of the linearly polarized emission, producing results in agreement with our experimental findings.

Furthermore, it is well known that for InGaAs/GaAs QD shape anisotropies have a significant effect upon the optical polarization properties of the QDs [27]. The experimental observation of high DOLPs raises the question of how strongly shape anisotropies affect the optical properties of self-assembled *a*-plane InGaN/GaN QDs, and thus how reliable these QDs might be in generating highly polarized photons. Hence, we also provide the first investigation of the effect of dot anisotropy on the DOLP of *a*-plane InGaN QDs, and couple our findings to insights from statistically significant experimental studies.

The manuscript is organized as follows. In Sec. 2, the growth methodology, the optical characterization, and the theoretical framework are explained briefly. The results of analyses are presented in Sec. 3, where we start with theoretical studies followed by experimental investigations. Finally, we summarize our results in Sec. 4.

## 2. Method

*2.1 MOCVD growth of the QD sample and fabrication of nanopillars*

The *a*-plane InGaN QD sample was grown by metal-organic vapor phase epitaxy in a 6 × 2 in. Thomas Swan close-coupled showerhead reactor on an *r*-plane sapphire substrate using trimethylgallium, trimethylindium, and ammonia as precursors. The *a*-plane GaN template was prepared using a silicon nitride ($SiN_x$) interlayer method to control the defect density [28]. After the deposition of a low temperature GaN nucleation layer at 500 °C, a 1 µm thick seed GaN layer was grown at 1050 °C and a V/III ratio of 60. A single $SiN_x$ interlayer was grown at 900 °C using a silane ($SiH_4$) flow rate at 0.2 µmol min$^{-1}$. A three-dimensional growth step was carried out at a pressure of 300 Torr using a V/III ratio of 1900, which was followed by a two-dimensional coalescence step using a V/III ratio of 60 at 100 Torr. The as-grown *a*-plane GaN template, utilizing the $SiN_x$ interlayer, has a dislocation density of $4 \times 10^9$ cm$^{-2}$ and a basal plane stacking fault density of $2.6 \times 10^5$ cm$^{-1}$ [28,29].

The InGaN QDs were grown using a modified droplet epitaxy method [30]. A 2.5 nm thick InGaN epilayer was grown at 695 °C and 300 Torr, which was immediately annealed in $N_2$ atmosphere for 30 s at the growth temperature. This annealing process results in the formation of In/Ga metallic droplets, which will re-react with ammonia during the GaN capping to form InGaN QDs (an initial ~ 10 nm GaN cap was grown at the InGaN growth conditions, and another ~15 nm GaN was grown at 1050 °C using $H_2$ as the carrier gas). The QDs were positioned in the centre of a 50 nm thick intrinsic GaN layer (unintentionally doped GaN – uid GaN), which was clad by ~ 600 nm of Si-doped GaN (dopant concentration ~ $3 \times 10^{18}$ cm$^{-3}$) at the bottom and ~ 200 nm of Mg doped GaN (dopant concentration ~ $3 \times 10^{19}$ cm$^{-3}$) on the top. An atomic force microscopy (AFM) image of an uncapped sample following the growth and anneal of the InGaN layer and a schematic of the full sample structure are shown in figure 1(a) and 1(b) respectively.

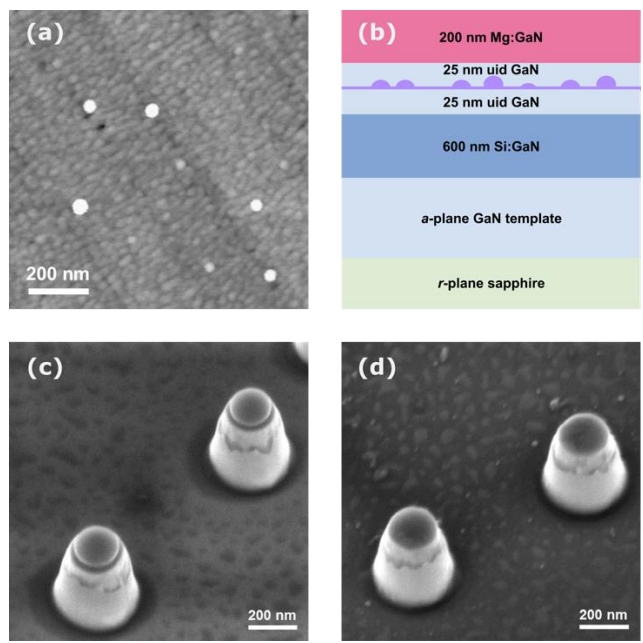

**Figure 1.** (a) AFM image of an uncapped QD sample grown on *a*-plane GaN template, with *z*-scale = 5 nm. (b) Schematic material structure for the nanopillar p-i-n sample with InGaN QDs. Also shown are SEM images (30 °tilted view) of nanopillars (c) before and (d) after the removal of the residual silica nanospheres.

Silica nanospheres (d ≈ 180 nm, fabricated in our laboratory [31]) were dispersed in ethanol and deposited onto the sample by drop casting. The nanospheres act as the etch mask for the nanopillars. The sample was subsequently dry-etched in an Oxford Instruments PlasmaPro 100 inductively coupled plasma etch system to a depth of ~ 350 nm using a gas mixture of $Cl_2$ and Ar. The residual silica nanospheres were removed by ultra-sonication in acetone, followed by a buffered oxide etch. SEM images of the nanopillars before and after the removal of the residual silica nanospheres are shown in figures 1(c) and 1(d) respectively, where the diameter at the top of the nanopillar is ~ 180 nm. The tapered shape is formed naturally during the dry-etching process.

*2.2 Polarization-resolved Microphotoluminescence*

The sample was mounted in an AttoDRY800 closed-cycle cryogenic system that maintained a stable sample temperature of 5 K. As shown in figure 2, an 80 MHz Ti:Sapphire laser system generates 1 ps pulses at 800 nm, providing two-photon excitation for the sample. It has been previously established that under multiphoton excitation, structures with higher degrees of quantum confinement have greater relative absorption cross-section [32]. As such, two-photon excitation produces better signal-to-background ratios for our sample which contains both QDs and a fragmented QW.

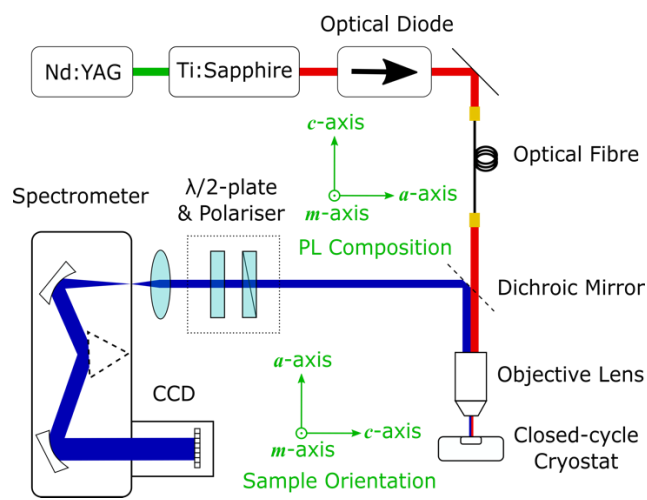

**Figure 2.** Schematic of polarization-resolved micro-photoluminescence (µ-PL) setup. Also shown are the sample orientation and consequent PL composition before passing through the polarizer in the plane of view.

A single-mode fibre with a 4 μm core spatially filters and transmits the excitation laser pulses onto a collimating lens, which then passes the beam to a 100× NIR objective with a numerical aperture of 0.5, producing a ~ 1 μm laser spot for sample excitation. A power density of 2 MW cm$^{-2}$ is chosen so that the QDs have the strongest emission before saturation takes place. The PL from the sample is then collected by the same objective and directed to an Andor Shamrock 500i half-metre spectrograph with variable slit size and a grating of 1200 lines mm$^{-1}$. An Andor iDus 420 Si-based charge-coupled device (CCD) is thermoelectrically cooled to – 50 °C for low-noise detection, enabling μ-PL measurements with a spectral resolution of ~ 38 pm.

For polarization-resolved μ-PL, a rotatable polarizer cube and half-wave plate were introduced in the optical collection path. The transmission axis of the polarizer was set to 0° on the rotation mount, and aligned parallel to the *c*-axis component of the sample PL. Hence, a polarizer rotation of *θ* degrees would correspond to the angle away from the *c*-axis in the growth plane (cf. figure 2).

Ensuring high accuracy in polarization-resolved μ-PL spectroscopy is experimentally challenging in a number of ways, including random sample drift, PL intensity fluctuations, and difficulty in measurement when approaching the weaker polarized component. In order to ensure the precision of measurement as far as possible, three precautions have been taken during the course of measurement: (1) The half-wave plate was rotated by – *θ*/2 for a polarizer angle of *θ*, thus maintaining a constant axis of PL polarization and correcting a ~ 10% difference in polarization-sensitive detection of the system. (2) The slit of the spectrometer was opened wide enough to account for slight optical alignment changes caused by the rotation of the polarizer cube, so that PL intensity remained constant at angles 180° apart. (3) To counteract any sample position drift, a reference measurement of the maximum intensity was immediately made after the data at each *θ* was taken and μ-PL results were normalised accordingly during analysis.

*2.3 Theoretical Simulations*

To address the electronic and optical properties of *a*-polar InGaN/GaN QDs from a theoretical perspective, we applied a flexible **k p** model, as described in detail elsewhere [17,33–35]. To study the DOLP, band mixing effects in the valence band are of central importance. As such, a six-band Hamiltonian to describe the hole states and a single-band effective mass approximation for electrons is used. In the following study, the *x*-axis of our coordinate system is parallel to the crystal *c*-axis. A schematic illustration of the coordinate system is given in the inset of figure 3(a).

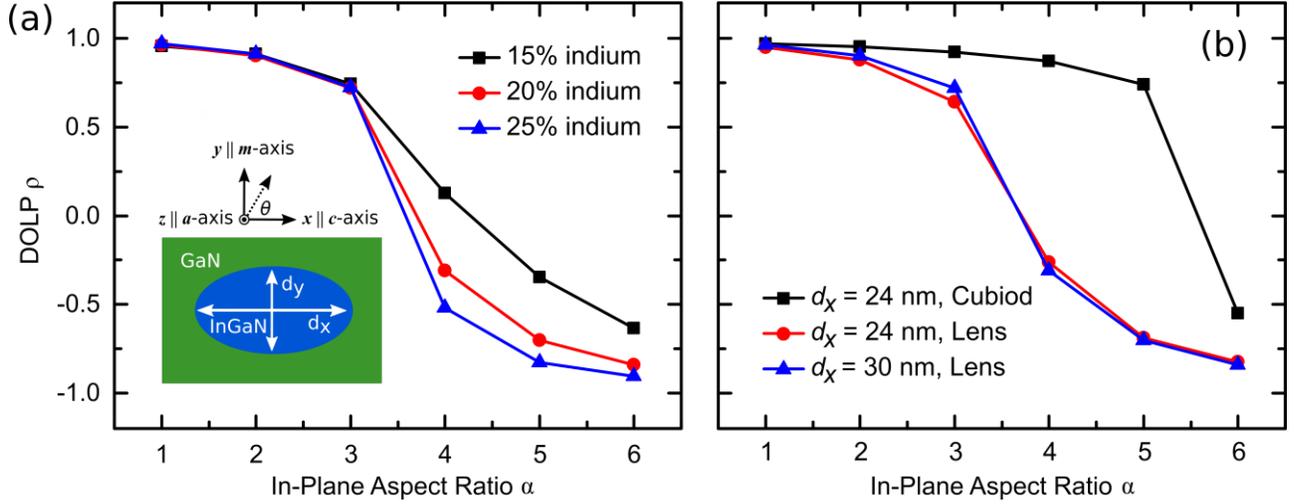

**Figure 3.** DOLP $\rho$ as a function of the in-plane aspect ratio $\alpha$ and indium content of a lens-shaped *a*-plane InGaN/GaN QD. The inset shows the coordinate system, the definition of $\theta$ as the in-plane angle away from the crystal *c*-axis (same as that defined in the experimental setup), and the definitions of $d_y$ and $d_x$ as the in-plane dimensions of a QD used in the theoretical investigation. The in-plane aspect ratio $\alpha$ is defined as $\alpha = d_x/d_y$.

## 3. Results and discussion

### 3.1 Theoretical calculations of DOLP

*3.1.1 DOLP of Symmetrical QD.* Knowledge about QD geometry and indium content, which is difficult to obtain experimentally, was required for calculation inputs. Previous studies on InGaN QD systems assumed lens-shaped geometries [36,37]. We followed this assumption, and based on earlier atomic force microscope results [30], chose a base diameter of 30 nm and a height of 2.5 nm as our starting point, with indium content varied between 15% and 25%. All calculations were performed on a $50 \times 50 \times 30$ nm$^3$ supercell with periodic boundary conditions. With these assumptions, the calculated ground state transition energies were in the range of 2.7 to 2.9 eV, close to typical experimental values [24,29,30,38–41] and those discussed below. Therefore, the chosen geometries and indium contents should give a reasonable description of the structures considered here.

The outcome of our **k·p** calculations for a dot with a circular-base predicted a hole ground state with 97% $|Y\rangle$-like and only 2% $|X\rangle$-like character, indicating that the emitted light should be polarized perpendicular to the $c$-axis (along the $m$-axis). The intensities measured perpendicular and parallel to the wurtzite $c$-axis, $I_\perp$ and $I_\parallel$, should correspond to $|Y\rangle$- and $|X\rangle$-like orbital contributions respectively. According to the DOLP formula, $\rho = (I_\perp - I_\parallel)/(I_\perp + I_\parallel)$, an $a$-plane InGaN QD with a circular base is hence predicted to have a $\rho$ of 0.96. This result is in stark contrast with theoretical studies of $c$-plane InGaN QDs, where $\rho$ is effectively 0 for dots with no in-plane anisotropies [15,16].

*3.1.2 Effect of in-plane anisotropy and indium content.* However, it is important to note that self-assembled QDs are unlikely to be perfectly symmetrical. Results on both InGaAs QDs [27] and $c$-plane InGaN QDs [15,16] reveal that dot shape anisotropy plays an important role and affects their properties quite significantly due to band mixing effects. As anisotropy affects the degree of quantum confinement, which causes shifts in the $|X\rangle$-, $|Y\rangle$-, and $|Z\rangle$-like states, it would be important to study how significant this effect is in our QDs by modifying the dot geometry in the simulation. We define the in-plane aspect ratio $\alpha$ as $\alpha = d_x/d_y$, where $d_x$ and $d_y$ are the in-plane dimensions of the dot base along $x$- and $y$-directions respectively. Here, for a circular base, $d_x = d_y = 30$ nm, and $\alpha = 1$. A schematic illustration of the situation is given in the inset of figure 3(a).

A smaller dimension, and thus stronger confinement effects, along the $x$-direction will further increase the DOLP as it approaches unity. This is attributed to the lower effective mass of the $|X\rangle$-like states along this direction and an increase in the energetic separation of $|Y\rangle$- and $|X\rangle$-like states. Correspondingly, the more interesting situation is what happens if the dot geometry is modified along the $y$-direction. Such an anisotropy should affect states with a high $|Y\rangle$-like orbital contribution more strongly, again due to the fact that $|Y\rangle$-like ($|X\rangle$-like) states exhibit a low (high) effective mass along the $y$-direction [42]. Please note that the coordinate system used in this work is different from that used in Ref. 42. More details on coordinate system used here can be found in Ref. 35. Hence, calculations were performed at $d_y = 30,15,7,6$ and 5 nm, as the QD base becomes more elliptical. The results for $\rho$ as a function of $\alpha$ and indium content are shown in figure 3(a). We see that $\rho$ is almost constant for $\alpha$ values between 1 and 2, and decreases only slightly for $\alpha = 3$, independent of indium content. It is important to note that $\alpha = 3$ already presents a significant "deformation" of the dot ($d_x = 30$, $d_y = 10$). Consequently, the calculations show that in $a$-plane InGaN QDs, the DOLP $\rho$ is very *insensitive* against shape anisotropies and indium composition changes.

*3.1.3 Effect of QD size and geometry.* To strengthen the argument that the DOLP is extremely robust against shape anisotropies, we have also varied the geometry and size of the dot. Firstly, to study the impact of the QD size on the DOLP, we have kept the geometry to be lens-shaped but reduced the in-plane dimensions of the system. Here, for the symmetric dot, $d_x = d_y = 24$ nm. To consider the same range of in-plane aspect ratios $\alpha$, $d_y$ has been varied between 24 and 4 nm. The results of this study are shown in figure 3(b) (red circles), revealing that the change in the QD in-plane dimensions affects the DOLP only very slightly.

To further extend this analysis, we have also investigated the influence of the QD geometry on the DOLP. To this end we have drastically changed the QD geometry from a lens-shaped dot to a cuboid. The length and width of the cuboid is assumed to be of 24 nm with a height of 2.5 nm. Figure 3(b) shows the data (black squares) of the DOLP as a function of $\alpha$ for the cuboidal dot. As one can infer from this study, in comparison with the lens shaped systems, the reduction of the DOLP happens at higher $\alpha$ values ($\alpha = 5$). Consequently, the DOLP in cuboidal shaped dots would even be more robust against shape anisotropies.

As such, our theoretical studies predict high experimental DOLP values, with very small variations caused by QD shape anisotropies, indium content fluctuations, size and geometry differences. To validate the simulation results and gain further insights from the experiments, we perform optical characterization of $a$-plane InGaN/GaN QDs.

*3.2 Optical characterization of a-plane InGaN QDs*

3.2.1 *Evidence of m-oriented highly polarized emission.* Using the method detailed in Sec. 2.2, the emission intensities of a single QD measured at 10° intervals are shown in the inset of figure 4. The data were fitted with $I(\theta) = I_\perp cos^2(\theta - \varphi) + I_\parallel sin^2(\theta - \varphi)$, where $I_\perp$, $I_\parallel$, and $\theta$ follow the same definitions explained previously, and $\varphi$ is the angle at which the maximum of function occurs. As both cross-polarized components should exhibit Malus' law type intensity variation, two sinusoidal fits with a phase difference of 90° are used in the equation. The sinusoidal intensity change thus demonstrates that the QD emission is polarized. Furthermore, the $\varphi$ is found to be 92° ± 1°, indicating that the maximum intensity occurs at $\theta$ = 92° and 272°, and indeed corresponds to the direction perpendicular to the *c*-axis. The small difference of ~ 2° is attributed to the error of sample placement in the cryostat.

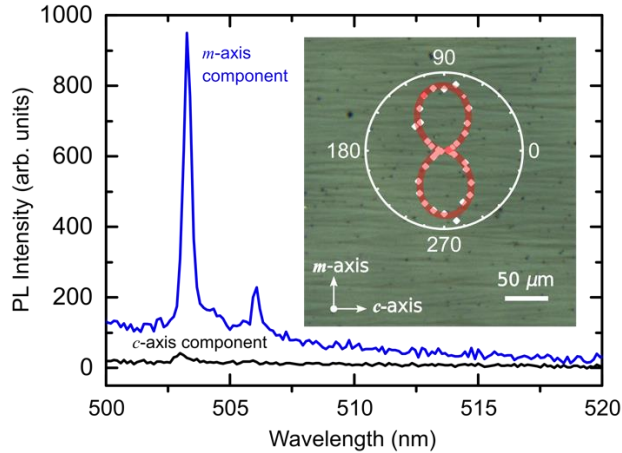

**Figure 4.** μ-PL of the two cross-polarized components for a QD at 503 nm (~ 2.46 eV). The inset demonstrates sinusoidal intensity variation of the QD emission and axis of polarization with respect to striations on the sample surface.

According to illustration in figure 2 and the definition of $\theta$, the polarization axis can hence be determined to be along the *m*-axis, the same as the simulation predicts. To further quantify how polarized the emission is, we invoked the same DOLP formula described previously and calculated a $\rho$ of 0.92 for this QD, again showing good agreement with the theoretical results for $\alpha$ values between 1 and 2.

Additionally, the intensity variation plot is shown on top of an optical microscope image of the sample. The *a*-plane sample surface exhibits striated features along the *c*-axis [28], further demonstrating the orthogonality between the polarization axis and the crystal *c*-direction. As such, the PL polarized parallel to the *c*-axis should correspond to the weaker polarized component of QD emission, instead of the stronger one as previously reported [24].

*3.2.2 Statistical studies of QD DOLP and comparison to QW.* In order to achieve statistical significance for this finding, we investigated the *absolute values* of DOLP of 180 QDs individually. The distribution of $|\rho|$ for these QDs is shown in figure 5(a). The high mean $|\rho|$ of 0.90 provides direct evidence that non-polar *a*-plane InGaN QDs are strongly polarized photon emitters. Furthermore, due to the modified droplet epitaxy growth routine [30], the *a*-plane QDs studied in this work were formed on top of fragmented QWs (fQWs). In the μ-PL experiments, we observe sharp emission features from the QDs together with the underlying fQW emission over the same spectral range, such as that in figure 4. As such, we are able to not only measure the DOLP for the QDs, but also for the underlying *a*-plane fQWs. For each of the 180 *a*-plane QDs we investigated in the main text, the $|\rho|$ of their respective fQWs are shown in figure 5(b).

All $|\rho|$ data of QDs fall in the range between 0.60 to 1.00, in contrast to those of QWs between 0 and 0.80. The average $|\rho|$ of 0.46 for the QWs, which is in agreement with the latest literature findings [23], is also much less than the average of 0.90 for the QDs. It could be speculated that the actual separation of different hole states in *a*-plane QDs is greater than that in QWs due to the additional degrees of quantum confinement, thereby resulting in higher $|\rho|$ values. From a fundamental perspective, this is interesting because out of *a*-plane nanostructures, only *a*-plane QDs possess such high DOLPs. From a technology viewpoint, coupled with the ability to achieve single-photon emission, *a*-plane QDs are well suited for the generation of polarized single-photons and useful for the development of quantum information applications.

The spread of the values figure 5(a) is attributed to variations of QD size, shape, and indium content. There are 23 QDs for which the weaker polarized component cannot be detected at all, giving a DOLP of unity. As explained in the theoretical simulation, this could be caused by a compression of the QD geometry along the *c*-direction. Those with DOLP values lower than average could be attributed to a compression along the *m*-axis (cf. figure 3(a) and 3(b)). Furthermore, the calculated standard deviation is only 0.08, which signifies very small fluctuations around the high average DOLP. As such, the robustness of DOLP against changes in size, shape anisotropy and indium content have been demonstrated both theoretically and experimentally. The data in figure 5(a) (5(b)) indicate that there are no discernible correlations between QD (QW) energy and DOLP, thus *a*-plane InGaN QDs emit highly polarized photons across all attainable wavelengths.

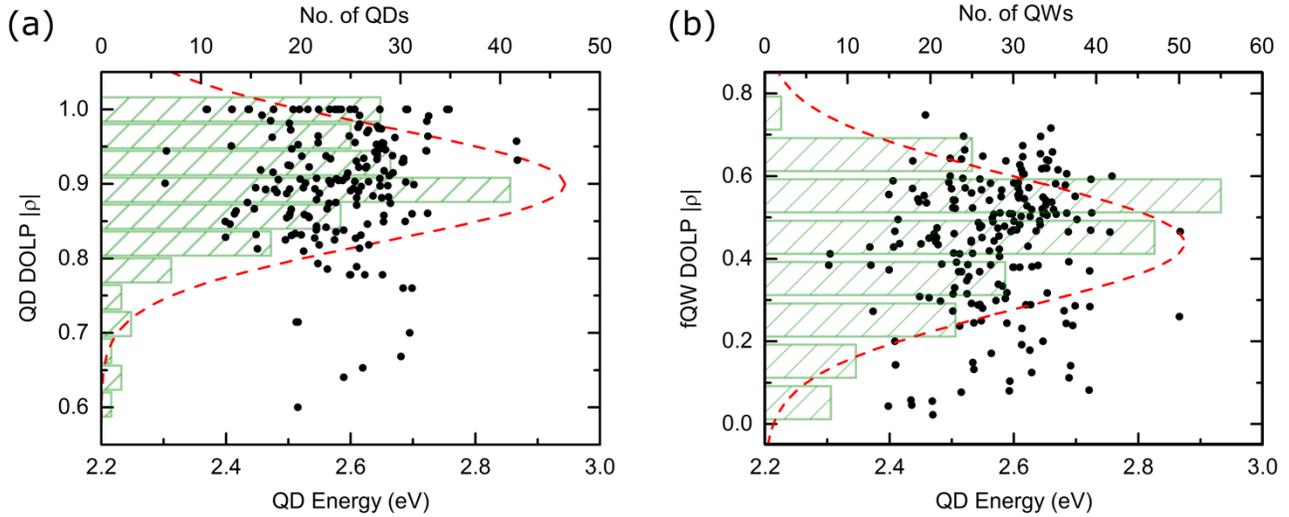

**Figure 5.** 180 *a*-plane InGaN/GaN (a) QD and (b) underlying QW DOLP variation with emission energy. The statistical distributions have been fitted with a Gaussian profile. The means and standard deviations of the Gaussian distributions are (a) $0.90 \pm 0.08$ and (b) $0.46 \pm 0.14$ respectively.

*3.2.3 Statistical studies of polarization axis.* The alignment of the polarization axis was also studied during the polarization measurement. In order to avoid selection bias, the polarizer was initially set to 45°. For each QD emission observed in the PL spectrum, the polarizer was then rotated from 0° to 90°, to determine the angle at which the maximum and minimum intensities occurred. 91% of the QDs exhibit polarization aligned along the *m*-axis (within a 10° error), which agrees with the simulation results. However, there are 9% of the studied QDs polarized along the *c*-axis. In our theoretical framework, this would mean that the hole ground state is predominantly $|X\rangle$-like. From figure 3(a), one can infer that $\rho$ drops significantly for $\alpha > 3$, and more so with higher indium content, as the confinement effects due to increased band offsets become stronger, making asymmetries in the QD geometry more prominent. Producing a predominant $|X\rangle$-like hole ground state would hence require an extreme deformation of the QD ($d_x$ = 30 nm, $d_y$ = 6 nm) and very high indium content, which is an unlikely configuration. Recently, atomistic calculations, including random alloy fluctuations, on non-polar InGaN/GaN QWs have shown that 15% to 20% of the different microscopic configurations considered can result in such emission properties [43]. Therefore, it is likely that random alloy fluctuations inside non-polar InGaN QDs can lead to similar effects, explaining that 9% of the dots emit light polarized along the *c*-axis. As such, the previous report [24] is an investigation of these 9% QDs, which is a special case to the general behaviour of *a*-plane InGaN QDs.

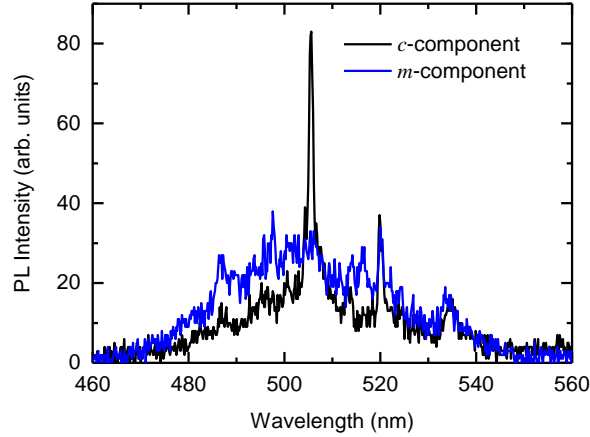

**Figure 6.** μ-PL of a QD (~ 505 nm) and its underlying fQW. The QD is polarized along the *c*-direction, while the fQW is polarized along the *m*-direction, corresponding to one of the 9% special cases in which *a*-plane QDs are polarized parallel to the crystal *c*-axis.

An example of an *a*-plane QD polarized along the *c*-direction, as one of the 9% cases, is shown in figure 6. The QD at ~ 505 nm has its maximum intensity at $\theta = 0°$, corresponding to a polarization axis along the crystal *c*-direction. However, the underlying *a*-plane fQW still behaves as expected, and has its maximum intensity at $\theta = 90°$. The QD has a DOLP of unity, and the QW has a DOLP of 0.27, both of which are in agreement with the statistics in figure 5(a) and 5(b) respectively. As such, we show evidence of cross-polarized QD and QW emissions from our *a*-plane InGaN QD sample. Due to the suppression of the QW background in these 9% cases, we are able to achieve higher dot-to-background ratios for the QDs with a polarization axis along the *c*-axis, which are beneficial for the development of purer single-photon sources with $g^{(2)}(0)$ values closer to 0. Overall, our experimental and theoretical data show that comparing to QDs fabricated with *c*-plane nitrides, arsenides, or other semiconductor materials, *a*-plane InGaN/GaN QDs are ideal candidates to achieve efficient linearly polarized photon emission with consistently high DOLPs and a deterministic polarization axis.

## 4. Conclusion

In conclusion, this work provides a statistically significant investigation of the polarization properties of *a*-plane InGaN QDs with high experimental precision and rigorous theoretical foundation. Cross-polarized photon emissions from 180 QDs produce an average DOLP as high as 0.90, with a small standard deviation of only 0.08. Theoretical **k p** simulations have also demonstrated similarly high DOLP for these QDs, which is almost insensitive to small size differences, shape anisotropies, and indium content changes. The polarization axis has been theoretically and experimentally determined to be mostly along the crystallographic *m*-axis direction, with a small (9%) minority along the *c*-axis possibly due to random alloy fluctuations. It is also important to note that the studied *a*-plane InGaN QD systems out-perform *a*-plane QW systems in terms of the DOLP. The statistically high DOLP, fixed polarization axis, and insensitivity to size, shape, and indium contents offer *a*-plane InGaN QDs unique advantages in polarization control comparing to their *c*-plane nitride counterparts, and to QDs based on other materials, such as InAs or CdSe. In summary, non-polar *a*-plane InGaN QD systems are ideal candidates for the development of on-demand polarized single-photon sources and related applications.

## Acknowledgements

This research was supported by the UK Engineering and Physical Sciences Research Council (EPSRC) Grants EP/M012379/1 and EP/M011682/1, and the Science Foundation Ireland (Project No. 13/SIRG/2210).

## Author contributions

T.W. and S.K.P contributed equally to this work. T.W. performed the optical experiments with the assistance of T.J.P, and S.K.P. conducted the theoretical studies. T.Z. prepared the sample, and with the assistance of M.A, T.B, and T.D, performed the nanopillar etching. R.A.O. supervised the sample preparation, S.S. supervised the theoretical investigation, and R.A.T. supervised the optical experiments.

## Data access

All data presented in figure 3–6 are available free of charge at the DOI: 10.5287/bodleian:g7x51vxeJ.